\documentclass[1p,times,procedia]{elsarticle}
\usepackage{nupha_ecrc}
\usepackage{float}

\volume{00}

\firstpage{1}

\journalname{Nuclear Physics A}

\runauth{U.~Heinz and J.~Liu}

\jid{nupha}

\jnltitlelogo{Nuclear Physics A}

\usepackage{amssymb}

\usepackage{lineno}
\usepackage[colorlinks = true,
            linkcolor = blue,
            urlcolor  = blue,
            citecolor = blue,
            anchorcolor = blue]{hyperref}

\biboptions{comma,sort&compress}

\usepackage[figuresright]{rotating}
\usepackage[dvipsnames]{xcolor} 

	\newcommand{\eqref}[1]{(\ref{#1})}

\begin{document}

\vspace*{-6mm}
\begin{frontmatter}



\dochead{}

\title{Pre-equilibrium dynamics and heavy-ion observables\tnoteref{1}} 
\tnotetext[1]{Work supported by the DOE, Office of Science, Office of Nuclear Physics under Award No. \rm{DE-SC0004286}. Computing resources provided by the 
\href{http://osc.edu/ark:/19495/f5s1ph73}{Ohio Supercomputer Center.}
Thanks to Chun Shen for fruitful discussions.}


\author{Ulrich Heinz and Jia Liu}

\address{Physics Department, The Ohio State University, Columbus, OH 43210, USA\\[-5ex]}

\begin{abstract}
To bracket the importance of the pre-equilibrium stage on relativistic heavy-ion collision observables, we compare simulations where it is modeled by either free-streaming partons or fluid dynamics. These cases implement the assumptions of extremely weak vs. extremely strong coupling in the initial collision stage. Accounting for flow generated in the pre-equilibrium stage, we study the sensitivity of radial, elliptic and triangular flow on the switching time when the hydrodynamic description becomes valid. Using the hybrid code {\sc iEBE-VISHNU} \cite{Shen:2014vra} we perform a multi-parameter search, constrained by particle ratios, integrated elliptic and triangular charged hadron flow, the mean transverse momenta of pions, kaons and protons, and the second moment $\langle p_T^2\rangle$ of the proton transverse momentum spectrum, to identify optimized values for the switching time $\tau_s$ from pre-equilibrium to hydrodynamics, the specific shear viscosity $\eta/s$, the normalization factor of the temperature-dependent specific bulk viscosity $(\zeta/s)(T)$, and the switching temperature $T_\mathrm{sw}$ from viscous hydrodynamics to the hadron cascade {\sc UrQMD}. With the optimized parameters, we predict and compare with experiment the $p_T$-distributions of $\pi$, $K$, $p$, $\Lambda$, $\Xi$ and $\Omega$ yields and their elliptic flow coefficients, focusing specifically on the mass-ordering of the elliptic flow for protons and Lambda hyperons which is incorrectly described by {\sc VISHNU} without pre-equilibrium flow.       
\end{abstract}


\end{frontmatter}



\section{Introduction}
\label{sec1}
%
Relativistic viscous hydrodynamics has become the workhorse of dynamical modeling of ultra-relativistic heavy-ion collisions. However, hydrodynamics does not become valid until the medium has reached a certain degree of local momentum isotropization \cite{Heinz:2015gka}. In an inhomogeneous system, collective flow (i.e. space-momentum correlations) begins, however, to develop already before hydrodynamics becomes valid. The hydrodynamic stage thus starts with a non-vanishing pre-equilibrium flow \cite{Kolb:2002ve,Broniowski:2008qk,Schenke:2012wb,vanderSchee:2013pia,Romatschke:2015dha}. In \cite{Liu:2015nwa} we therefore performed a systematic study of pre-equilibrium flow effects on heavy-ion collision observables. We here summarize the main results of this study and add a few recent results that go beyond the work reported in \cite{Liu:2015nwa}.

\vspace*{-3mm}
\section{The model}
\label{sec2}
\vspace*{-2mm}
%
We assume longitudinal boost-invariance and model the approach towards local thermal equilibrium in a heavy-ion collision very simply \cite{Liu:2015nwa} as a pre-equilibrium stage of noninteracting free-streaming massless partonic degrees of freedom, separated by a switching time $\tau_s$ from a strongly coupled quark-gluon plasma (QGP) stage in which frequent collisions isotropize the local momentum distribution sufficiently quickly that it can be described by viscous relativistic fluid dynamics. The variable switching time $\tau_s$ parametrizes the duration of the thermalization process. For massless degrees of freedom the evolution of the energy-momentum tensor during the free-streaming stage is independent of the initial transverse momentum distribution as long as it starts out locally azimuthally symmetric, and can be solved analytically \cite{Liu:2015nwa}. 

At $\tau_s$ we Landau-match the analytically evolved pre-equilibrium energy-momentum tensor to viscous hydrodynamic form \cite{Liu:2015nwa}. Space-momentum correlations established by the free-streaming dynamics in the pre-equilibrium stage manifest themselves as non-zero initial values for the hydrodynamic flow velocity profiles, giving rise to non-zero initial radial and anisotropic flows. Local momentum anisotropies resulting from the free-streaming evolution of a spatially inhomogeneous initial density profile generate nonzero initial values for the shear stress. Matching the traceless pre-equilibrium energy-momentum tensor of noninteracting massless degrees of freedom to a non-conformal, lattice QCD based equation of state \cite{Bazavov:2014pvz} at $\tau_s$ generates a non-zero initial bulk viscous pressure field in the hydrodynamic fluid. All of these initial fields are then further evolved with the second-order viscous relativistic fluid dynamics code {\sc VISH2+1}. 

After hadronization of the QGP at a pseudocritical temperature $T_\mathrm{c}{\,\simeq\,}155$\,MeV (determined by the equation of state \cite{Bazavov:2014pvz}), the medium constituents are color neutral hadrons whose interactions are weaker than the gluon-mediated interactions in the earlier color-deconfined QGP, causing a rapid growth of their mean free paths and a subsequent breakdown of the fluid dynamic framework. We therefore switch from {\sc VISH2+1} to a microscopic description, using the hadron cascade code {\sc UrQMD}, on an isothermal switching surface of temperature $T_\mathrm{sw}$ which we here consider as an unknown parameter to be optimized by comparison with experimental data.

Two additional important parameters affecting the evolution of collective flow and the $p_T$ distributions of the finally emitted hadrons are the shear and bulk viscosities during the liquid QGP stage. For simplicity, we assume the QGP specific shear viscosity $\eta/s$ to be a temperature-independent, adjustable constant. For the specific bulk viscosity $\zeta/s$, which is expected to develop a strong peak due to critical scattering close to the quark-hadron phase transition, we adopt the temperature-dependent parametrization given in\cite{Denicol:2009am} (which features a peak at $T_\mathrm{peak}=180$\,MeV), but consider its normalization as a freely tunable parameter. In practice, peak values of $\zeta/s$ exceeding 1 (corresponding to bulk viscosity normalization factors ${\,>\,}3$) cause the (negative) bulk viscous pressure to become very large near $T_\mathrm{peak}$, leading to negative total pressures and mechanical instability of the medium against cavitation. These physical instabilities also eventually render the hydrodynamic code numerically unstable. 

\vspace*{-3mm}
\section{Results}
\label{sec3}       
\vspace*{-2mm}  
%
We found in \cite{Liu:2015nwa} that an extended pre-equilibrium stage increases the final radial flow, leading to flatter $p_T$-spectra, while leaving the integrated charged hadron $v_2$ and $v_3$ almost unchanged unless $\tau_s$ significantly exceeds 2\,fm/$c$. Bulk viscosity inhibits radial flow build-up, so the constraint from the measured $p_T$ distributions causes a strong positive correlation between $\tau_s$ and the bulk viscosity normalization factor. The experimental charged hadron elliptic flow constraint anticorrelates bulk and shear viscosity; at low $p_T$, the bulk viscous correction $\delta f_\mathrm{bulk}$ at $T_\mathrm{sw}$ acts as an effective positive chemical potential for massive hadrons \cite{Dusling:2011fd}, causing an anticorrelation between the bulk viscosity normalization factor and the chemical decoupling temperature $T_\mathrm{sw}$.

\begin{figure}[H]
\centering
\includegraphics[width=0.7\linewidth]{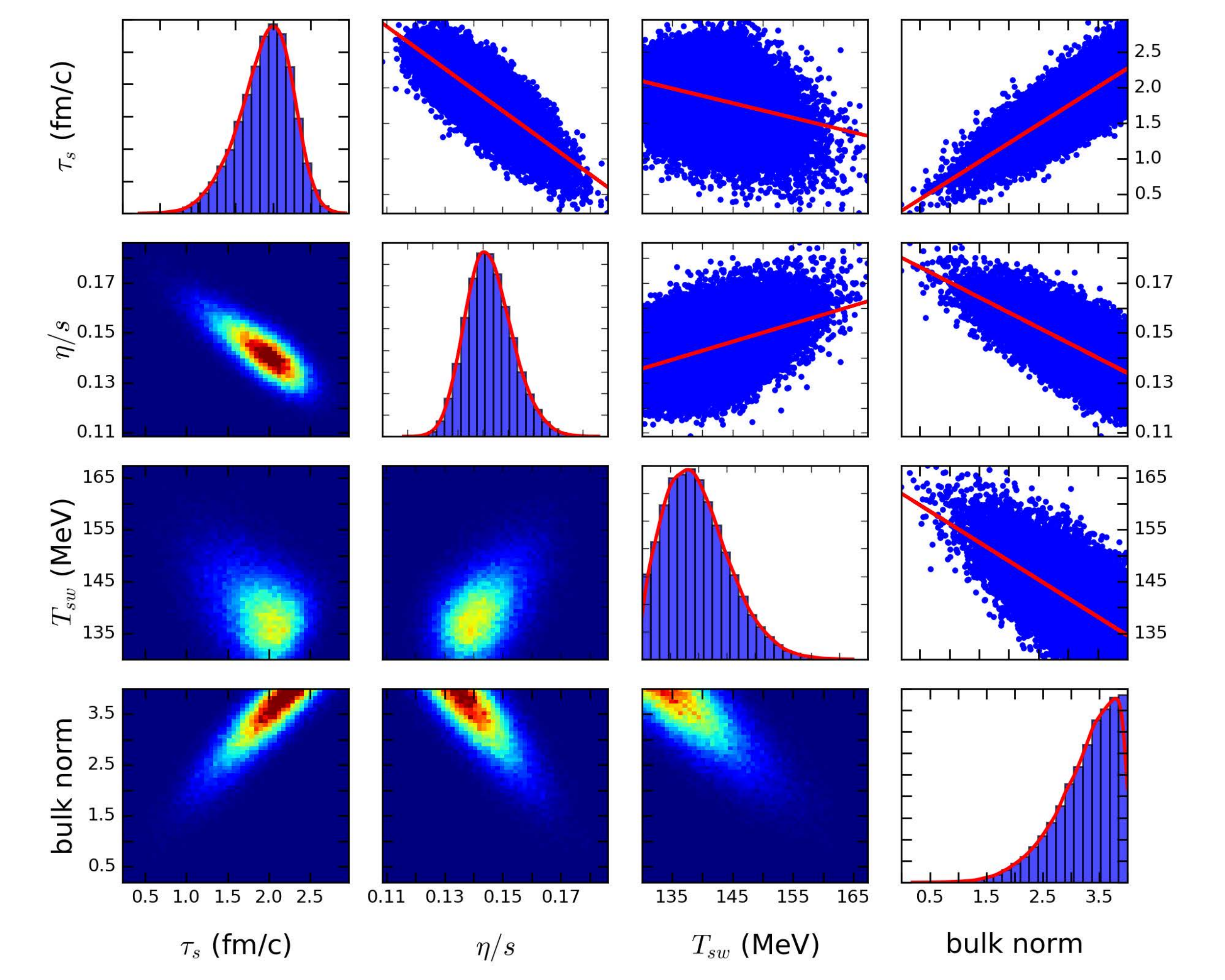}\\
\includegraphics[width=0.32\linewidth]{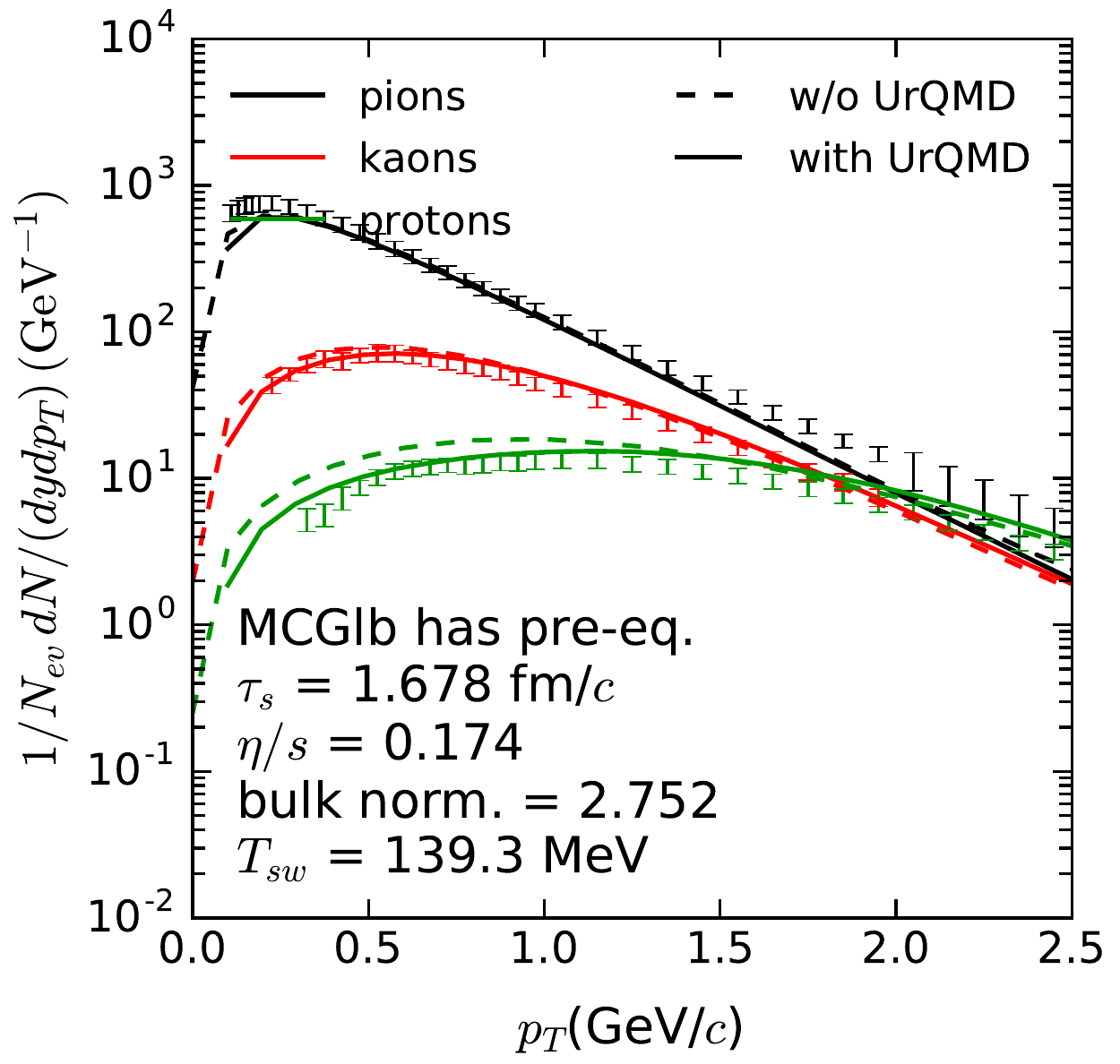}		
\includegraphics[width=0.32\linewidth]{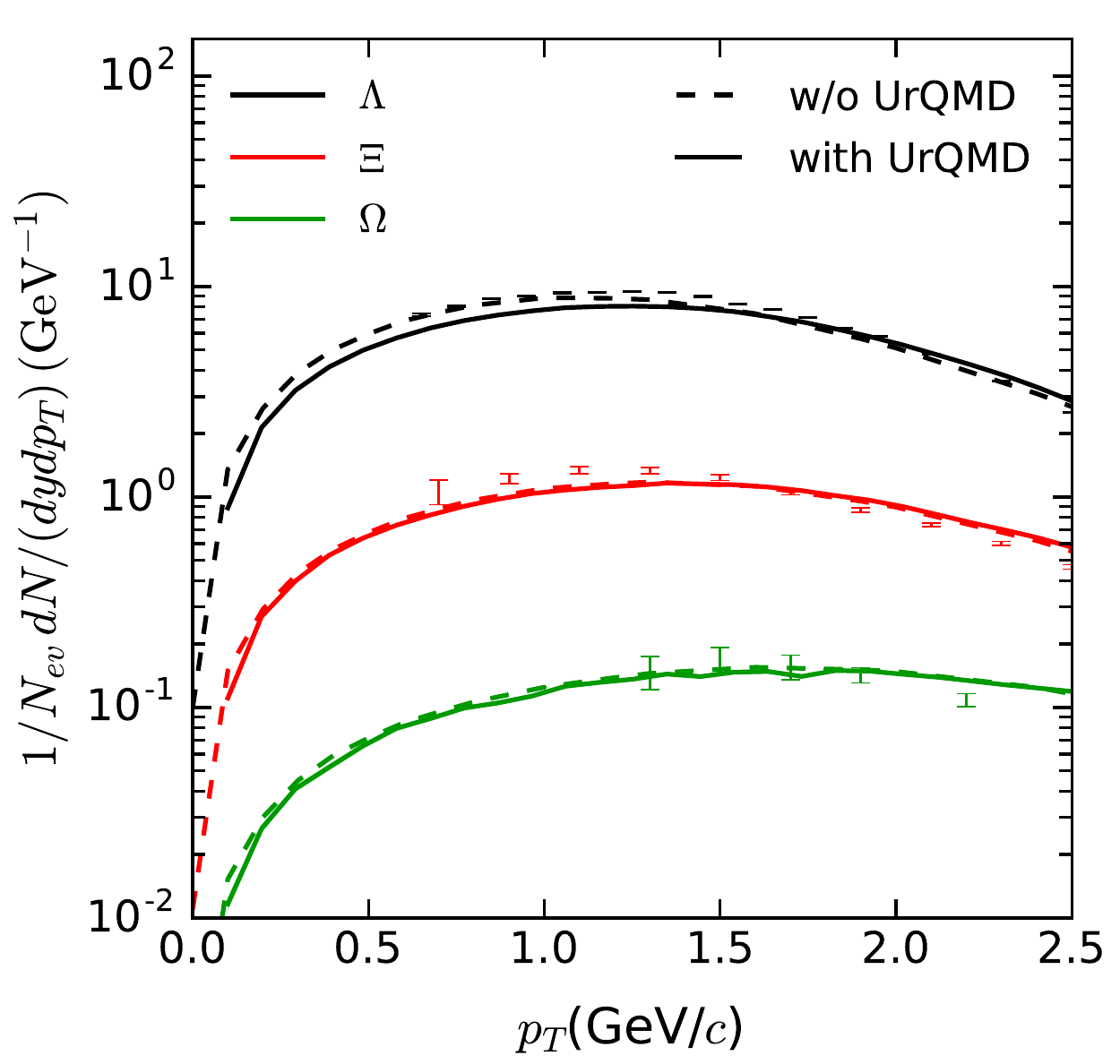}		
\includegraphics[width=0.32\linewidth]{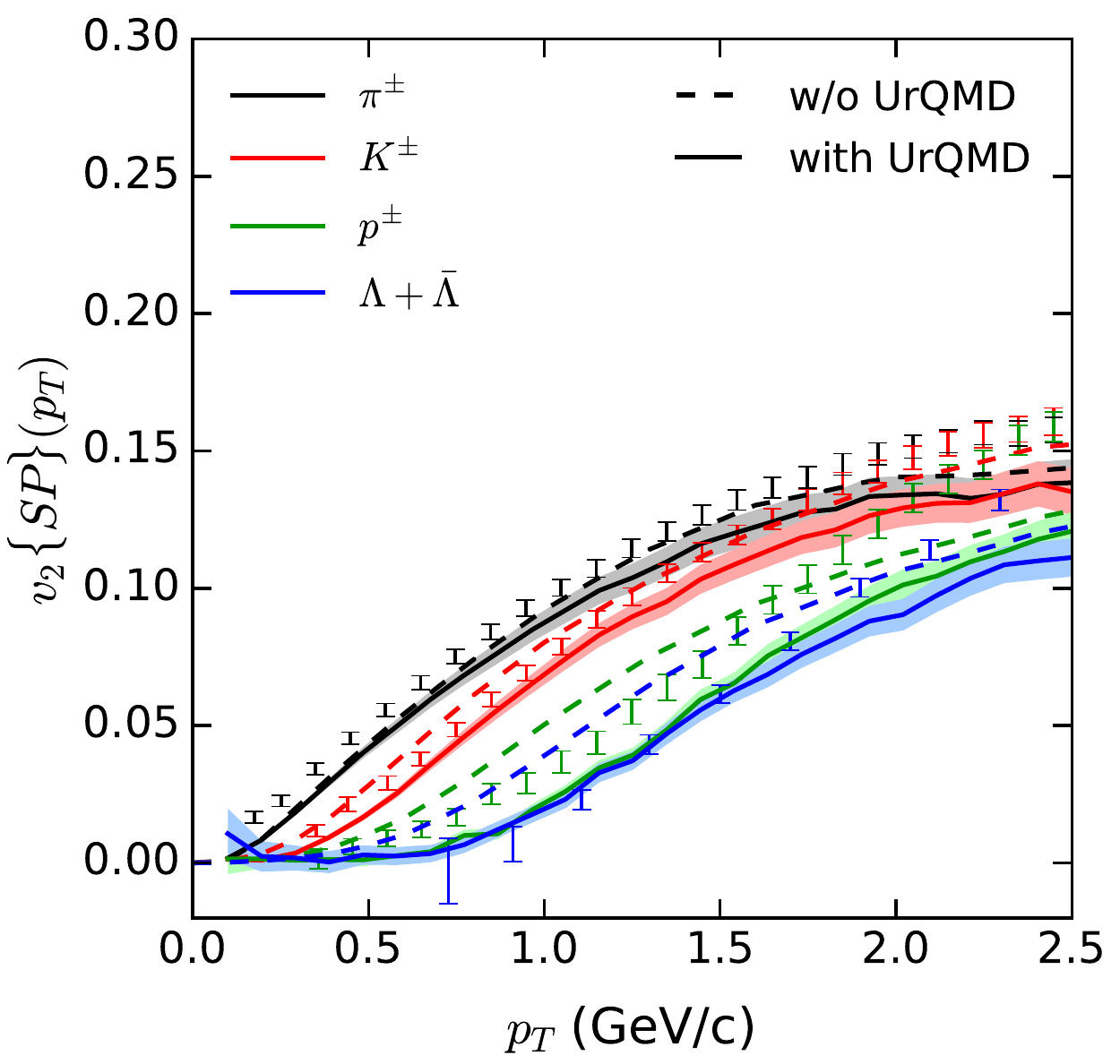}
\vspace*{-2mm}		
\caption{{\sl Upper panel:} Likelihood distributions for each of the four fit parameters are shown on the diagonal, correlations between them in upper right triangle of plots, and 2-dimensional projections of the likelihood distributions for pairs of parameters (with dark red indicating the most likely regions) in the lower triangle of plots. {\sl Lower panels:} $p_T$ distributions for pions, kaons and protons (left), $\Lambda$, $\Xi$ and $\Omega$ (middle), and the differential elliptic flows for $\pi$, $K$, $p$ and $\Lambda$ (right) from 400 hydrodynamic simulations with fluctuating initial conditions, each oversampled 400 times with {\sc UrQMD} to obtain sufficient particle statistics. Solid lines: full {\sc VISHNU} runs. Dashed lines: pure hydro runs with immediate freeze-out at $T_\mathrm{sw}$, without subsequent hadronic rescattering using {\sc UrQMD}. Experimental data from ALICE \cite{Preghenella:2012eu}.    
\label{Fig1}}
\end{figure}

These tendencies are illustrated in the top panel of Fig.~\ref{Fig1} for which we ran {\sc VISHNU} simulations with smooth ensemble-averaged MC-Glauber initial conditions for 10-20\% central Pb-Pb collisions at 2.76\,$A$\,GeV for 1000 different quadruplets of the parameters listed in Table~\ref{T1}, calculated the experimental observables listed in the same table (chosen for their insensitivity to initial-state event-by-event fluctuations), determined the $\chi^2$ for each quadruplet by comparing them with the experimental values, and then reconstructed the posterior likelihood distributions in the 4-dimensional parameter space using MCMC sampling \cite{madai_manual}. For the set with the lowest $\chi^2$, listed in the lower left panel of Fig.~\ref{Fig1}, we then ran 400 event-by-event {\sc VISHNU} simulations with fluctuating initial conditions, each oversampled 400 times in the {\sc UrQMD} stage for sufficient particle statistics, to predict the full transverse momentum distributions and their elliptic flows $v_2(p_T)$ for the specific hadron species shown in the bottom row of Fig.~\ref{Fig1}. 

\begin{table}[!h]
\begin{tabular}{lcc}
\hline \hline 
$\langle v_2^{\mathrm{ch}}\rangle$ \ \cite{Aad:2013xma}  &$0.0782 \pm 0.0019$ \\ 
$\langle v_3^{\mathrm{ch}}\rangle$ \ \cite{Aad:2013xma} &$0.0316 \pm 0.0008$  \\ 
$\langle p_T\rangle_{\pi^+}$ (GeV/$c$) \ \cite{Abelev:2013vea}& $0.542\pm 0.018$  \\ 
$\langle p_T\rangle_{K^+}$   (GeV/$c$) \  \cite{Abelev:2013vea}& $0.825 \pm 0.028$   \\ 
$\langle p_T\rangle_{p}$    (GeV/$c$) \  \cite{Abelev:2013vea}& $1.311\pm 0.043$ \\  
$\langle p_T^2\rangle_{p}$ (GeV$^2$/$c^2$) \  \cite{Abelev:2013vea}& $2.085\pm 0.070$ \\
$(dN_{\pi^+}/dy)/(dN_{K^+}/dy)$ \cite{Abelev:2013vea}\hspace*{-5mm}    &        $6.691  \pm  0.670$ \\
$(dN_{\pi^+}/dy)/(dN_{p}/dy)$     \cite{Abelev:2013vea}       &        $21.667 \pm 2.292$\\
$(dN_{\pi^+}/dy)/(dN_{\Lambda}/dy)$  \cite{Abelev:2013vea}&       $26.765 \pm 3.639$ \\
\hline\hline
\end{tabular}
\begin{tabular}{lccccc}
\hline \hline 
parameter           &  best         & mean   & 95\% C.I.\\ \hline \hline
$\tau_s$ (fm/$c$)   & 2.233        & 1.889   & 1.187 - 2.591          \\
$\eta/s$            & 0.135        & 0.143   & 0.124 - 0.161          \\
$T_\mathrm{sw}$ (MeV)      & 133.4        & 134.0   & 128.5 - 151.1        \\
bulk norm.          & 3.998        & 3.277   & N/A       \\\hline\hline
\end{tabular}
\caption{{\sl Left:} Measured values of nine hadronic observables from 2.76\,$A$\,TeV 10--20\% central 
              Pb+Pb collisions used to constrain the model parameters listed on the right. $\langle p_T\rangle$
              and $\langle p_T^2\rangle$ are truncated means calculated from the data tables for the transverse
              momentum distributions listed in \cite{Abelev:2013vea}.
              {\sl Right:}      
              Best-fit parameters and 95\% confidence intervals (C.I.) using the results from Markoff Chain 
              Monte Carlo sampling of the posterior parameter distribution shown in Fig.~\ref{Fig1} (top). 
              \label{T1}}
\end{table}
\vspace*{-3mm}

To illustrate the effect of the microscopically simulated late hadronic rescattering stage on these observables we also added dashed lines illustrating their state at the end of the fluid stage at $T_\mathrm{sw}$. Protons and $\Lambda$s experience a significant radial boost from {\sc UrQMD}, pushing their yields and elliptic flows to larger transverse momenta. Since $\Lambda$s in {\sc UrQMD} scatter with reduced cross sections relative to those of protons, this shift towards larger $p_T$ is weaker for $\Lambda$s than for protons; as a result, the hydrodynamically predicted elliptic flow mass ordering between protons and $\Lambda$s at $T_\mathrm{sw}$, clearly visible in the dashed lines in the lower right panel of Fig.~\ref{Fig1} and also in the experimental data \cite{Preghenella:2012eu}, is basically eliminated after hadronic rescattering, although not inverted as previously observed in the {\sc VISHNU} model without pre-equilibrium dynamics \cite{Zhu:2015dfa}. A shorter lifetime of the hadronic rescattering stage could help to keep this from happening.

%
%
%




\bibliographystyle{elsarticle-num}



%

\end{document}